\documentstyle[preprint,aps]{revtex}
\begin{document} 
\thispagestyle{empty} 
\begin{flushright}
UA/NPPS-7-1999
\end{flushright}
\begin{center}
{\large{\bf STATISTICAL BOOTSTRAP ANALYSIS OF \\ 
$\bf S+Ag$ INTERACTION AT 200AGeV:\\ 
EVIDENCE OF A PHASE BEYOND THE HADRONIC ONE? \\}} 
\vspace{2cm} 
{\large A. S. Kapoyannis, C. N. Ktorides and A. D. Panagiotou}\\ 
\smallskip 
{\it University of Athens, Division of Nuclear and Particle Physics,\\ 
GR-15771 Athina, Hellas}\\ 
\end{center}
\vspace{3cm}
\begin{abstract}
A generalized Strangeness-incorporating Statistical Bootstrap Model (SSBM) is
constructed so as to include 
indepedent fugacities for up and down quarks. Such an extension is crucial
for the confrontation of multiparticle data emerging from heavy ion
collisions, wherein isospin symmetry is not satisfied. Two constraints, in
addition to the presence of a critical surface which sets the boundaries of the
hadronic world, enter the extended model. An analysis pertaining to
produced particle multiplicities and ratios is performed
for the $S+Ag$ interaction at 200 GeV/nucleon. The resulting
evaluation, concerning the location of the source of the produced system,
is slightly in favor the source being outside the hadronic domain.
\end{abstract}
\vspace{1cm}
\newpage
\setcounter{page}{1}

{\Large \bf 1. Introduction}

\vspace{0.1cm}
Multiparticle production in high energy collisions is a subject of intense 
research interest, whose history goes almost as far back as that of the strong 
interaction itself. Indeed, it registers as one of the key features entering 
the analysis of collision processes, involving the strong force, at both the 
experimental and the theoretical fronts. With specific reference to relativistic 
heavy ion collisions, the task of accounting for the produced 
multiparticle system is by far the most important issue to consider for 
extracting information of physical interest. 

A notably successful theoretical approach, through which experimentally 
observed particle multiplicities have been confronted, is based on the idea of 
thermalization. Within such a context, one views the multiparticle system, 
emerging from a given high energy collision, as being comprised of a large 
enough number of particles to be describable in terms of a thermodynamical set 
of variables. Relevant, standard treatments appearing in the literature [1-5] 
adopt an ``Ideal Hadron Gas'' (IHG) scheme, wherein any notion of interaction 
is totally absent\footnote{Only the repulsive form of interaction has been
introduced in some cases through a hard sphere model.}.

The fact that such analyses, ranging from $e^+e^-$ to $A+A$ collisions, 
produce very satisfactory results simply verifies, a posteriori, that the 
thermalization assumption is justifiable. Beyond this realization, however, no 
fundamental insight and/or information is gained with respect to the {\it 
dynamics} operating during the process, which produced the multiparticle system 
in the first place. Given, in particular, that the object of true interest, in 
the case of relativistic heavy ion collisions, is whether the original thermal 
source of the multiparticle system is traceable to a region that belongs, or not, 
to the hadronic phase, an IHG-type of analysis renders itself totally inadequate. 
Clearly, only if interactions are taken into consideration does it become 
relevant to ask whether or not a change of phase has taken place during the 
dynamical development of the system. 

In a recent series of papers [6-8], we have pursued a line of investigation 
which, on the one hand, approaches the study of multiparticle systems from the 
hadronic side (just as the IHG case) while, on the other, incorporates the 
effects of interactions in a self-consistent way. We are referring to the 
employment of a scheme, known as Statistical Bootstrap Model (SBM), which was 
originally introduced by Hagedorn [9-11], much before QCD was conceived and was 
subsequently developed via notable contributions by a number of authors. 
Excellent reviews articles on the SBM can be found in Refs [12].

The crucial feature of the SBM is that it adopts a statistical-thermodynamical
mode of description, which admits interactions among its
relativistic constituent particles via a bootstrap logic. According to the SBM, 
the constitution of the system is viewed at different levels of organization 
(fireballs) with each given level being generated as a result of interactions 
operating at the preceeding one. The remarkable feature of the SBM is that the 
so-called bootstrap equation (BE), which results from the aformentioned 
reasoning, defines a critical surface in the space of the thermodynamical 
variables, which sets an upper bound to the world of hadrons and implies, under 
precisely specified conditions, the existence of a new phase of matter beyond. 

Let us briefly review this construction while giving, at the same time, an 
overview account of the bootstrap scheme itself. We start by displaying the
generic bootstrap equation, whose final form reads 
\begin{equation} 
\varphi(T,\{\lambda\})=2G(T,\{\lambda\})-\exp(T,\{\lambda\})+1\;, 
\end{equation} 
where $\varphi$ is the so-called {\it input function}, since its 
specification involves an input from all observable hadrons and $G$ 
incorporates, via the bootstrap logic, the mass-spectrum of the system given 
in terms of fireballs of increasing complexity. Note that, $\varphi$ and $G$
as functions depend on a thermodynamic set of variables (temperature and 
fugacities). 
 
The key feature of the BE is that it exhibits a square root branch point 
at 
\begin{equation} 
\varphi(T,\{\lambda\})=\ln 4-1\;, 
\end{equation} 
which defines a critical surface in the space of thermodynamical variables that
sets the limits of the hadronic phase, in the
sense that eq. (1) does not posses physically meaningful solutions beyond this 
surface. This is not to say that the BE is thermodynamically consistent with the 
existence of a different phase on the other side. In this connection, the 
deciding factor is the form of the so-called spectrum function $\rho(m^2)$ 
entering the definition of $G$ and, in particular, the way it factorizes into a 
kinematical and a dynamical part (see following section). 
 
The final ingredient of the SBM is the employment of a grand canonical 
partition function $Z(V,T,\{\lambda\})$, which accounts for thermodynamical 
properties. In combination with the BE, it furnishes a thermal description of a 
system comprised of relativistic entities (hadrons/
\linebreak fireballs)
{\it interacting with each other}. 
 
Recognizing the importance of the role played by the quantum number of 
strangeness in providing possible signals for a presumed QGP phase, we have 
extended the SBM by introducing a fugacity variable for strangeness into the scheme 
[6,7]\footnote{The latter was not included in the original SBM as the main 
preocupation at the time referred to nuclear matter.}. We shall be referring to 
the resulting extended construction as the ``SSBM''. 
 
Imposing the condition $<S>=0$ we proceeded to study thermodynamical 
properties of the SSBM. Central emphasis was placed on the choice of the 
spectrum function, in order to acheive an acceptable thermodynamical
description, consistent with the existence of a phase beyond the hadronic one. The 
end result is encoded into the following relation expressing the partition 
function in terms of the ``bootstrap function'' $G$ [6,7]
 
\begin{equation} 
\ln Z(V,T,\{\lambda\})=\frac{VT^3}{4\pi^3 H_0} 
\int_0^T \frac{1}{y^5}G(y,\{\lambda\})dy\;, 
\end{equation} 
where $H_0\equiv\frac{2}{(2\pi)^3 4B}$ with $B$ the MIT bag constant. 
 
In combination with the critical surface condition furnished by the BE, 
one is able to relate the critical temperature $T_0$ at vanishing chemical 
potential with $B$. This occurence provides a direct connection between
QCD-inspired phenomenology and critical temperature for the hadronic state of 
matter. Our numerical studies have been based on the choice $T_0=183$ MeV, which 
corresponds to the maximum acceptable value for $B^{1/4}$, namely 235 MeV. Such 
a choice is consistent with the strangeness chemical potential $\mu_s$ remaining 
positive definite throughout the hadronic phase while maximally extending  
the region of the hadronic phase and thereby rendering our 
appraisal of the proximity of the source of the multiparticle system  to the 
critical surface (or beyond) as conservative as possible. 
 
Subsequently, we generalized the SSBM [8] by introducing a further 
``fugacity'' variable $\gamma_s$, which allows for {\it partial} strangeness chemical 
equilibrium. This extension of the model enables us to confront the data with an 
open perspective on strange particle production, as we let the observed particle 
multiplicities and ratios determine whether strangeness saturation has taken 
place or not. 
 
In [8] we also conducted a systematic study of multiparticle states (particle 
multiplicities and particle ratios) produced in $S+S$ as well as in $p+\bar{p}$ 
collisions at CERN (experiments NA35 and UA5, respectively), the latter considered more as a test case. Our results  
yield an excellent account of particle multiplicities and ratios (equally good, 
if not slightly better than IHG results [5,14,15]). More importantly,  
we have identified a region in the 
space of thermodynamical parameters where the source of the produced 
multiparticle state is expected to lie and appraised its location with
respect to the limiting surface of the hadronic phase.

For the $S+S$ interaction we summarize
the highlights of our findings as follows:
 
(a) The quality of our results were similar to that given by IHG analysis. 
This further justifies the thermalization hypothesis. 
 
(b) Almost full saturation of strangeness was observed, which accounts for an 
enhanced production of strange particles relative to non-strange ones. 
 
(c) The source of the multiparticle system was found to lie just outside the 
limits of the hadronic phase, as established by the SSBM. 
 
(d) An excess of pions (SSBM/experimental = 0.73) is observed, which is not
fully compatible with the theoretical prediction of a purely hadronic phase. 
At the same time, entropy considerations also give SSBM/QGP $=0.71-0.78$,
pointing, together with (c), to a source being in the doorway of a deconfined
phase.

These findings strongly suggest that in the $S+S$ interaction at 200 
GeV/nucleon the thermalized, strangeness-saturated source of the multiparticle 
system has exceeded the hadronic sector and has entered the lower limits of the 
QGP phase.

Now, the $S+S$ colliding system is symmetric under isospin transformations,
hence consistent with the simplification $\lambda_u=\lambda_d$ adopted in our
previous work on the SSBM. In the present paper we shall further extend the
model so as to accomodate isospin non-symmetric systems. Such a step will
enable us to confront multiparticle data for the $S+Ag$ collision experiment
(NA35), at CERN. As we shall see, this further extension imposes a new
constraint on the system which relates charge and baryon numbers. It follows
that the SSBM extension we shall be discussing amounts, at the hadronic level,
to introducing a fugacity variable pertaining to total charge.

The presentation of the new extension of the SSBM, accomodating
isospin non-symmetric systems, will be accomplished in Section 2.
The profile of the relevant construction, accentuated by the presence of
the critical surface in the space of thermodynamical parameters as well as
the two surfaces resulting by the imposition of the two physical constraints,
will be discussed.

Our confrontation of the data (particle multiplicities and ratios) for the
$S+Ag$ experiment (NA35 at CERN) is presented in Section 3.
Our concluding remarks are
made in Section 4. Two appendices are devoted to corresponding discussion of
a more specialized nature. In Appendix A we establish that the value for the
critical temperature for zero chemical potentials corresponds to a
maximum on the critical surface. Appendix B discusses the subtle points
involved in the minimization of the $\chi^2$-variable given the presence of
constraints and the critical surface beyond which
the SSBM has no analytical validity.

\vspace{2cm}
{\Large \bf 2. Isospin non-symmetric SSBM}

\vspace{0.1cm}
In this section we shall realize the construction of a maximally extended
SSBM, accomodating both partial strangeness saturation and isospin
asymmetry. In this way we shall be in position to perform thermal analyses
pertaining to (non)strange particle production in nucleus-nucleus collisions
in which the total number of participating protons differs from that
of neutrons.

\vspace{0.5cm}
{\large \bf 2a. Preliminaries}

\vspace{0.1cm}
The set of variables in terms of which the initial quantification of the
bootstrap scheme is accomplished naturally associates itself with input
particle (and fireball) attributes. These are number densities and
four-momenta pertaining to particle/fireball species. As hinted to in the
introduction the situation we wish to consider in this paper involves, the
following number densities: Baryon number $b$, net strangeness $s$, overall
strangeness $|s|$ and ``net charge'' $q$. The employed sequence respects
``historical'' order in the following sense. In the original SBM only $b$
enters, the SSBM construction of Ref. [6,7] includes $s$(=strangeness minus
anti-strangeness number) while the extension of Ref [8] has added
$|s|$(=strangeness plus anti-strangeness number) to the list.

Our present effort amounts to a further extension of the SSBM through which
we incorporate a ``net charge'' number density into the bootstrap scheme. To
quantify our considerations regarding this new variable let us focus on the
initial states entering a nucleus-nucleus collision process and consider the
ratio $\frac{N_p^{in}}{N_n^{in}}$, where $N_p^{in}$($N_n^{in}$) denotes the
total number of protons(neutrons) participating in the collision. Suppose
this ratio is equal to unity. It then, follows that
\begin{equation}
\frac{N_p^{in}}{N_n^{in}}=\frac{<Q>^{in}}{<B>^{in}-<Q>^{in}}=1\;,
\end{equation}
where $<Q>^{in}$ and $<B>^{in}$ are the incoming total charge and baryon
numbers, respectively. Equivalently, the above condition reads
\begin{equation}
<B>=2<Q>\;.
\end{equation}
By introducing a ``net charge'' particle density into the bootstrap scheme we
declare our intention to confront $A+A$ collision processes which do not, a
priori, respect the condition given by (5). It is not hard to see that the
latter corresponds to an isospin non-symmetric system, at least as far as
its initial (incoming) composition is concerned.

With reference to (5) the quantification of isospin asymmetry can be
parametrized as follows
\begin{equation}
<B>=\beta 2<Q>\;,
\end{equation}
where
\begin{equation}
\beta=\frac{N_p^{in}+N_n^{in}}{2N_p^{in}}.
\end{equation}
Our actual preoccupation, of course, is with the description of the produced,
final states. Accordingly, we shall eventually impose (6) as a {\it
constraint} on the system.

We close this general exposition with a brief discussion of the particular
version of the SSBM we have adopted throughout our work as far as the
issue of dynamics vs kinematics is concerned.
Generically speaking, the SSBM construction involves a mass spectrum function $\tau$ whose
dependence is on the set of variables $\{p^2,b,s,|s|,q\}$. 
A kinematic factor $\tilde{B}(p^2)$ enters the equation (see the BE in the
following subsection), the specific choice of which
classifies different versions of bootstrap models, according to asymptotic
behaviour, as the fireball mass goes to infinity. Our specific commitment to
the form $\tilde{B}(p^2)$ has been discussed at great lenght in Refs. [6,7]. We have
argued that there are desicive physical advantages in favor of the choice
\begin{equation}
\tilde{B}(p^2)=B(p^2)=\frac{2V^{\mu}p_{\mu}}{(2\pi)^3}\;,
\end{equation}
where $V^{\mu}$ is the (boosted) four-volume associated with a given
particle/fireball and $p_{\mu}$ the corresponding four-momentum. The two
four-vectors being parallel to each other imply a relation of the form
\begin{equation}
V_{\mu}=\frac{V}{m} p_{\mu}\;,
\end{equation}
$V$ being the rest frame volume. We, therefore, have
\begin{equation}
B(p^2)\rightarrow B(m^2)=\frac{2Vm}{(2\pi)^3}\;.
\end{equation}

The mass mass spectrum aquires the asymptotic form
\begin{equation}
\tilde{\tau}(m^2,\{\lambda\})\stackrel{m\rightarrow\infty} 
{\longrightarrow} 
C'(\{\lambda\})m^{-1-\alpha} \exp [m/T^*(\{\lambda\})]\;\;\;. 
\end{equation}

The above relations determine the version ($\alpha=4$) of the bootstrap model we have
found to be physically relevant. It should be pointed out that the bulk of
the work surrounding the bootstrap model, prior to the introduction of
strangeness, was based on the choice $\alpha=2$ [13].

\vspace{0.5cm}
{\large \bf 2b. Construction of the model}
   
The initial form of the BE reads
\[
\tilde{B}(p^2)\tilde{\tau}(p^2,b,q,s,|s|)=
\underbrace{g_{bqs|s|}\tilde{B}(p^2)\delta_0(p^2-m^2_{bqs|s|})}_
{input\;term}+\sum_{n=2}^{\infty} \frac{1}{n!}
\int \delta^4 (p-\sum_{i=1}^n p_i)\cdot
\]
\[
\cdot \sum_{b_i} \delta_K (b-\sum_{i=1}^n b_i)
\sum_{b_i} \delta_K (q-\sum_{i=1}^n q_i)
\sum_{s_i} \delta_K (s-\sum_{i=1}^n s_i)
\sum_{|s|_i} \delta_K (b-\sum_{i=1}^n |s|_i)
\]
\begin{equation}
\prod_{i=1}^n \tilde{B}(p^2_i)\tilde{\tau}(p^2_i,b_i,q_i,s_i,|s|_i)d^4p_i\;.
\end{equation}
The new feature, with respect to our previous extensions of the bootstrap
model, is the introduction of electric charge $Q$ as an additional
variable\footnote{In our thermodynamical context the electric charge will enter
as charge number density $q$.}.

Performing five Laplace transforms (one continuous and four discrete) leads
to the following replacement of variables
\begin{equation}
(p^2,b,s,|s|,q)\rightarrow(T,\lambda_B,\lambda_S,\lambda_{|S|},\lambda_Q)\;,
\end{equation}
where the $\lambda$'s represent fugacity variables corresponding to number
densities and $T$ is the temperature, as recorded in the center of mass frame.

As the final states are composed of hadrons, rather than just baryons, we
find it more convenient to pass from the original set of fugacities into one
given in terms of valence quark fugacities. The transcription is made
according to the relations
\begin{equation}
\lambda_B=\lambda_u \lambda_d^2,\;\lambda_Q=\lambda_u \lambda_d^{-1},\;
\lambda_{|S|}=\gamma_s,\;\lambda_S=\lambda_d \lambda_s^{-1}\;.
\end{equation}
The important implication of the above relations is that they facilitate a
thermodynamical description of the system in terms of (valence) quark
fugacities, thereby enabling us to accomodate the presence of any kind of
hadronic particle in the final system. Specifically, the form of the
functions $\varphi$ and $G$ entering the bootstrap scheme is given by
\begin{equation}
\varphi(T,\lambda_u,\lambda_d,\lambda_s,\gamma_s;H_0)=
2\pi H_0 T \sum_{\rm a}
\lambda_{\rm a}(\lambda_u,\lambda_d,\lambda_s,\gamma_s)
\sum_i g_{{\rm a}i}m_{{\rm a}i}^3 K_1 \left( \frac{m_{{\rm a}i}}{T}\right)
\end{equation}
and
\begin{equation}
G(T,\lambda_u,\lambda_d,\lambda_s,\gamma_s;H_0)=
2\pi H_0 T \int_0^{\infty} m^3 \tau_0(m^2,\lambda_u,\lambda_d,\lambda_s,
\gamma_s) K_1 (m/T) dm^2\;,
\end{equation}
where $K_1$ denotes the modified Bessel function of the second kind and
where the general form of the fugacities $\lambda_{\rm a}$, pertaining to the
{\it totality} of hadronic families, is
\begin{equation}
\lambda_{\rm a}(\{\lambda\})=\lambda_u^{n_u-n_{\bar{u}}}
\lambda_d^{n_d-n_{\bar{d}}} \lambda_s^{n_s-n_{\bar{s}}}
\gamma_s^{n_s+n_{\bar{s}}}\;,
\end{equation}
where $n_i$ is the number of the $i$ quarks contained in the hadron of the
${\rm a}$ family.
For the particular case of the fugacities of light unflavored mesons, one
can employ the parametrizacion 
$c_1(u\bar{u}+d\bar{d})+c_2s\bar{s}$, with $c_1+c_2=1$, see Ref [8],
whereupon the corresponding variables assume the form
\begin{equation}
\lambda_{\rm a}(\{\lambda\})=c_1+c_2\gamma_s^2\;.
\end{equation}

The bootstrap equation (1), written analytically for the case in hand, reads
\begin{equation}
\varphi(T,\lambda_u,\lambda_d,\lambda_s,\gamma_s)=
2G(T,\lambda_u,\lambda_d,\lambda_s,\gamma_s)-
\exp[G(T,\lambda_u,\lambda_d,\lambda_s,\gamma_s)]+1\;,
\end{equation}
while the critical surface is determined by (either one of) the relations
\begin{equation}
\varphi(T_{cr},\mu_{u\;cr},\mu_{d\;cr},\mu_{s\;cr},\gamma_{s\;cr};H_0)=
\ln4-1
\end{equation}
and
\begin{equation}
G(T_{cr},\mu_{u\;cr},\mu_{d\;cr},\mu_{s\;cr},\gamma_{s\;cr};H_0)=\ln2
\end{equation}
Clearly, the critical surface corresponds to a 4-dimensional surface immersed
in the space of the 5 thermodynamical variables.

The constant parameter $H_0$, related directly to the MIT-bag constant (see
remark following eq (3) and Refs [6,7]), can also be linked to the critical
temperature at vanishing chemical potentials by
\begin{equation}
\varphi(T_0,\mu_u =0,\mu_d =0,\mu_s =0,\gamma_s;H_0)=\ln4-1\;.
\end{equation}

Through this relation $H_0$ can be directly related to $T_0$, for a fixed
value of the ``fugacity'' $\gamma_s$. At the same time we demonstrate, in
Appendix A, that $T_0$ corresponds to the maximum value for the temperature
on the critical surface, irrespective of the value $\gamma_s$. In our
previous work this feature was simply assumed.

In order to acquire a concrete sense concerning the profile of the critical
surface we have conducted a number of numerical studies which are displayed
in Figs. 1-3. In these figures we present various sections of the critical
surface, having chosen $H_0$ such that $T_0=183$ MeV for $\gamma_s=1$.

Fig. 1 depicts projections of the critical surface on the $(\mu_u,T)$-plane
for three representative values of $\mu_d$ and three for $\mu_s$. One
observes that the critical surface ``shrinks'' (equivalently, ``narrows'') as
$\mu_s$ reaches higher positive values, starting from zero. This
``shrinkage'' is more pronounced in the vicinity of vanishing $\mu_u$.

Fig. 2 displays critical surface projections on the $(\mu_u,\mu_d)$-plane for
three different values of $T$ and $\mu_s$. One notices that the projections
are (approximately) symmetric with respect to the line $\mu_u=\mu_d$. A
second point is that the lowering of $\mu_s$ causes an expansion of the
region occupied by the hadronic phase in the $(\mu_u,\mu_d)$ plane.

Finally, Fig. 3 shows projections on the $(T,\mu_s)$ plane of the critical
surface for fixed values of $\mu_u$ and of $\mu_d$. We notice that for fixed
$\mu_d$ and $\mu_s$ the critical temperature falls with increasing (absolute)
values of $\mu_u$. The same holds true under the exchange
$\mu_u\leftrightarrow\mu_d$.

\vspace{0.5cm}
{\large \bf 2c. Imposition of constraints}

\vspace{0.1cm}
Given the constitution of the initial colliding states, we must impose the
constraints $<S>=0$ and $<B>-\beta2<Q>=0$ on the system as a whole. To this
end we must refer to the partition function for our chosen version of the
bootstrap scheme, as given by eq. (3). The constraints have the generic form
\begin{equation}
H_k(T,\{\lambda\})\equiv\int_0^T \frac{1}{y^5}
\frac{F_k(y,\{\lambda\})}{2-\exp[G(y,\{\lambda\})]}dy=0\;,\;k=1,2\;,
\end{equation}
with
\begin{equation}
F_1(y,\{\lambda\})=\lambda_s
\frac{\partial \varphi(y,\{\lambda\})}{\partial \lambda_s}
\end{equation}
for the imposition of $<S>=0$ and
\begin{equation}
F_2(y,\{\lambda\})=
\frac{1-4\beta}{3}\lambda_u
\frac{\partial \varphi(y,\{\lambda\})}{\partial \lambda_u}+
\frac{1+2\beta}{3}\lambda_d
\frac{\partial \varphi(y,\{\lambda\})}{\partial \lambda_d}\;,
\end{equation}
for securing the constraint $<B>-\beta2<Q>=0$.

These conditions constitute a system of two equations whose solution yields
a 3-dimensional hypersurface in the space of thermodynamical variables on
which the given system is constrained to exist. Let us denote this surface by $H_{ph}$, where ``$ph$'' stands
for physical. Clearly, the intersection between $H_{ph}$ and the critical
surface defines the limits of the hadronic world for the system with the
given constraints. This intersection comprises a two-dimensional surface
whose numerical study is presented in Figs. 6-8.

Figures 4 and 5 give corresponding perspectives of the profile of $H_{ph}$
whose basic aim is to display its
variation with $\beta$. We have considered the cases $\beta=1$
$(N_p^{in}=N_n^{in})$, $\beta=2$ $(N_p^{in}<N_n^{in})$ and $\beta=1/2$
$(N_p^{in}>N_n^{in})$\footnote{For purposes of comparison we have also drawn
corresponding projections for the IHG model.}. Fig. 4 depicts projections of
$H_{ph}$ in the $(T,\mu_s)$-plane for fixed values of $\lambda_u$ and
$\gamma_s$, while Fig. 5 shows corresponding projections in the $(\mu_u,\mu_d)$
plane. From the first figure we record the tendency of $\mu_s$ to increase
with $\beta$, for fixed values of $(T,\lambda_u,\gamma_s)$. From the second
we witness the (expected) behavior $\mu_u=\mu_d$ for $\beta=1$, $\mu_u<\mu_d$
for $\beta>1$ and $\mu_d<\mu_u$ for $\beta<1$.

Finally, in Figs. 6-8 we present results of numerical studies pertaining to
the intersection between $H_{ph}$ and the critical surface. In Fig. 6 the
2-dimensional intersection is projected on the $(\mu_u,T)$-plane, for our
three representative values of $\beta$. As one might expect, an increase of
$\beta$ induces a decrease of $\mu_{u\;cr}$ for constant temperature. Fig. 7
shows the corresponding projections on the $(\mu_u,\mu_d)$-plane exhibiting
similar connections between $\beta$-values and the relation among
$\mu_{u\;cr}$ and $\mu_{d\;cr}$. In Fig. 8 we consider projections in the
$(\mu_u,\mu_s)$-plane. Here we surmise that for fixed value of $\mu_{u\;cr}$
an upward move of $\beta$ with respect to 1 $(N_p^{in}<N_n^{in})$ induces
an increase in the (critical) chemical potential of the strange quark.

This concludes our discussion of the isospin non-symmetric SSBM. We shall
proceed, in the next section, to confront experimental data encoded
in the multiparticle system produced in $A+A$ collisions, in which we do not
have isospin symmetry.

\vspace{2cm}
{\Large \bf 3. Analysis of $\bf S+Ag$ data of NA35}

\vspace{0.1cm}
In this section we shall perform a data analysis referring to particle
multiplicities recorded in the NA35 $S+Ag$ experiment at 200 GeV/nucleon at
CERN. The method we shall use is similar to the one presented in [8].
The main differences are that our space is described by the
set of the six thermodynamical variables $(VT^3/4\pi^3,T,\{\lambda\})$, i.e.
one more variable is present and that the system is subject to two
constaints, namely $<S>=0$ and $<B>=2\beta<Q>$, instead of one.
The latter will be enforced via the introduction of corresponding Lagrange
multipliers.

The theoretical values of the thermodynamical parameters are adjusted via a
$\chi^2$-fit by minimizing the function
\[\chi^2(VT^3/4\pi^3,T,\{\lambda\},\{l\})= 
\sum_{i=1}^N\left[\frac{N_i^{exp}-N_i^{theory} 
(VT^3/4\pi^3,T,\{\lambda\})} 
{\sigma_i}\right]^2\] 
\begin{equation} 
+\sum_{k=1}^2 l_k H_k(T,\{\lambda\})\;\;. 
\end{equation}
where $l_k$ are Lagrange multipliers accompanying the corresponding
constraints as given by (23) and the $N_i^{theory}$ are given by

\begin{equation} 
N_i^{theory}=\left.\left(\lambda_i\frac 
{\partial \ln 
Z(VT^3/4\pi^3,T,\{\lambda\},\ldots,\lambda_i,\ldots)} 
{\partial \lambda_i}\right)\right|_{\ldots=\lambda_i=\ldots=1} \;\;. 
\end{equation}

The minimization of $\chi^2$ ammounts to solving the following system of
eight equations
\begin{equation} 
\frac{\partial \chi^2(x_1,\ldots,x_8)}{\partial x_i}=0 
\;\;(i=1,\ldots,8)\;, 
\end{equation} 
with $\{x_i\}=(VT^3/4\pi^3,T,\{\lambda\},\{l\})$.

An outline of the procedure involved in realizing a numerical solution of the
minimization problem has been given in Ref. 10. There, we have also discussed the
methodology by which we determine correction factors for Bose/Fermi
statistics. We shall not repeat the general argumentation here, nevertheless
we do present in Appendix B a discussion of some technical aspects involved
in the relevant procedure for the case in hand.

Turning our attention to the $S+Ag$ collision at an energy of 200
GeV/nucleon, using the methodology that has just been described, we set as
our first task to specify the value of the $\beta$-parameter appropriate
for the process under study. As far as the $^{32}S$ nucleus is concerned, the
single isotope with nucleon number 32 ($Z=16$) is employed, whereas for
silver there are two stable isotopes with nucleon numbers 107 and 109
($Z=47$) entering, respectively, a mixture composed of $51.84\%$ and
$48.16\%$ fractions. This accounts for an average nucleon number of 107.96.
It turns out that it makes little difference whether one assumes that all the
nucleons entering the $S+Ag$ system participate in the collision process, or
that the ``active'' part of the $Ag$ nucleus is determined by some
``realistically'' assumed geometrical configuration. For our numerical
applications we shall fix the value of $\beta$ at 1.10. The emerging results
are displayed in a series of Tables and Figures.

In table 1 we present adjusted sets of values for the thermodynamical
parameters with a corresponding estimation for $\chi^2 /dof$\footnote{To ensure
that the $\chi^2$ estimate is carried out without leaving the domain of
analyticity of the SSBM the numerical values in table 1 correspond to
$T_0=190$ MeV for $\gamma_s=1$.}.
The presented numbers correspond to evaluations where all particle
multiplicities are taken into account (1st row) and where, in turn, one of
the particle species is excepted. One notices a decive improvement when
$h^-$ (mostly pions) are excluded from the fit. This occurence makes
meaningful the separate treatment of the full multiplicity analysis
from the one(s) where pions are excluded.

The experimental data pertaining to particle multiplicities have been taken
from [16-20,14,5] and are entered in the first column of Table 2.
The second column gives the theoretical estimates of
populations based on the corressponding adjusted set of thermodynamical
parameters with all particle species included. The third column pertains to the adjusted set with the absence of
pions. The last column corresponds to the same situation but with
the critical surface pushed slightly outwards by setting $T_0=183.5$ MeV at
$\gamma_s=1$.

Table 3 exhibits the correction factors due to Bose/Fermi statistics
for each particle species. We have covered each of the three
cases entering the previous table: All particle species, exclusion of pions
with $T_0=183$ MeV at $\gamma_s=1$ and $T_0=183.5$ MeV at $\gamma_s=1$,
respectively.
Table 4 summarizes the adjustment of the thermodynamical parameters
according to the $\chi^2$ fit (along with the estimate for $\chi^2/dof$) for
each of the three aformentioned cases. Finally, Table 5 presents
particle ratios (used only for the case where all the multiplicities are
included), taken with respect to negative hadron population which has
the smallest experimental uncertainty (see Ref [7] for a relevant comment).

Pictorial representation of results with physical significance is given
in Figs. 9-12. In the first of these figures we display
bands, per particle ratio, in the ($\mu_u,T$)-plane with $\gamma_s$ fixed
at 0.67 (see Table 4). These bands are determined by the experimental
uncertainty per ratio. The bold solid line marks the boundary of the
hadronic world, beyond which the SSBM does not present analytic solutions.
All particle ratios are used (as per Table 5), i.e. pions have not been
excluded  in the plot. No overlap region of the various bands is observed,
nevertheless we {\it have} marked with a cross the center of a region of
``optimum overlap'' which lies inside the hadronic world.

Fig. 10 considers corresponding bands of particle populations. Since the
variable $VT^3/4\pi^3$ also enters our considerations we fix it according to
its adjusted value of 1.23 (see second column of Table 4). It should be
pointed out that the upper limit for the experimental $K_s^0$ population (17)
as well as the whole band of the negative hadron population (175-197)
correspond to fitted values for the thermodynamical parameters that are
outside the hadronic domain (bold solid line). The dotted line marks the
smallest value of $\chi^2$ on the boundary surface. We have
determined that $51.6\%$ of particle multiplicities are compatible with a
source {\it outside} the hadronic domain. A zoom around the vicinity of the
minimal $\chi^2$-value on the critical surface is presented in Fig. 11.

A comparison between measured and theoretically determined, according to our
$\chi^2$ fit, multiplicities is summarized in Fig. 12. Experimental points,
with error bars, are represented by heavy dots. Theoretically determined
points are marked according to the three cases studied throughout, i.e. all
particle species inclusion and exclusion of pions with $T_0$ set,
respectively, at 183 MeV and 183.5 MeV for $\gamma_s=1$. Once again we notice
a dramatic improvement of the fits when pions ($h^-$) are excluded. As we have
argued in Ref. 10, this occurence seems to signify an excess of pion
production, incompatible with a pure hadronic phase (SSBM/experimental=
$0.69\pm0.04$).

\vspace{2cm}
{\Large \bf 4. Conclusions}

In this work we have applied the SSBM to analyze experimental data from the
$S+Ag$ collision at 200 GeV per nucleon pertaining
to produced particle multiplicities ($4\pi$ projection) 
recorded by the NA35 collaboration at CERN. The quintessential aspect of the
model is that it accomodates interactions in a self-consistent way within the
framework of a thermal description of the (relativistic) multiparticle system.
Moreover, it designates a precicely defined boundary for its applicability, a
feature which plays a central role in the assesment of our results.
It should finally be reminded that for the construction of this bounbary we
have chosen the largest possible, physically meaningful values of $T_0(B)$
so as to enlarge the hadron gas domain and avoid over-optimistic
interpretations, regarding the location of the source with respect to this
boundary.

Our primary objective has been to locate the {\it source} of the multipaticle
system in the space of the relevant thermodynamical set of parameters. In this
connection, we have found that the data point towards a thermal source that
lies just outside the hadronic phase. The situation is not as pronounced as
in the previously  analysed case of $S+S$ collisions [8] at 200 GeV per
nucleon, where a much larger weight in favor of the source being outside the
hadronic boundaries was determined. The overall situation resulting
from our data analyses ($p+\bar{p}$, $S+S$, $S+Ag$) is depicted in Fig. 13.
One notices the proximity of the source location for the two nucleus-nucleus
collision proccess just beyond the hadron phase as well as the (expected)
placement of the source for the $p+\bar{p}$ collision well inside the
hadronic domain.

The experimental data give a substantial excess of pion (entropy)
production compared to theoretical predictions. This strongly
hints that the source of the emerging multiparticle system from the $S+Ag$
collision is in the doorway of the QGP phase. As already pointed
out in the Introduction, an estimate of the entropy associated with the
pionic component of the produced system for the $S+S$ collision 
gives [8] a theoretical to experimental ratio which is not compatible with
hadronic physics and necessitates a location of the source outside the hadronic
domain. A similar behavior persists in the present case as well.

A final result of interest was the observance of a
tendency towards strangeness saturation. This indicates that the source has
acheived almost full thermal and chemical equilibrium, as expected and
required for a phase transition to QGP.

We thus conclude that in the $S+S$ and $S+Ag$ interactions at 200 GeV/nucleon
we have witnessed for the first time the appearance of definite signals
linking these interactions with the QGP phase.

The fully extended SSBM is now in position to confront
multiparticle data emerging from any $A+A$ collision
experiment, including strange particles.
In this respect, the methodology can be applied to other ongoing
experiments, e.g. $Pb+Pb$, as data becomes available and, more importantly,
on the future ones from RHIC and LHC. On the theoretical side, it would be extremely
interesting to connect a scheme such as the SSBM coming from the hadronic side
to corresponding microscopic-oriented accounts of QGP physics [21].

\newpage
{\Large \bf Appendix A}

\vspace{0.1cm}
We shall show that the critical temperature value $T_0$, as defined in the
text, corresponds to its maximum value on the critical surface.

We start by re-expressing eq. (20) in the form
\begin{equation}
T=f(\{\lambda\},\gamma_s)\;,
\end{equation}
where we ignore critical value indications on each variable for notational
simplicity.

A maximum for $T$ corresponds to an extremum
\begin{equation}
\left.\frac{\partial f (\{\lambda\},\gamma_s)}{\partial \lambda_i}
\right|_{\textstyle \varphi} =0\;,\;\;\;i=1,\ldots,4\;.
\end{equation}

As long as one remains on the critical surface the above condition can be
easily transcribed to
\begin{equation}
\left.\frac{\partial f (\{\lambda\},\gamma_s)}{\partial \lambda_i}
\right|_{\textstyle \varphi}=
-\frac{\partial\varphi/\partial\lambda_i}{\partial\varphi/\partial T}=
-\frac{\lambda_i\partial\varphi/\partial\lambda_i}
{\lambda_i\partial\varphi/\partial T}\;,\;\;\;i=1,\ldots,4
\end{equation}
and since $\lambda_i\partial\varphi/\partial T \neq 0$ it must be that
\begin{equation}
\lambda_i\frac{\partial\varphi}{\partial\lambda_i}=0\;,\;\;\;i=1,\ldots,4\;.
\end{equation}
Now, for the hadronic fugacities we may write
\begin{equation}
\lambda_{\rm a}=c_1+\lambda_b\{\lambda\}\gamma_s^{N_{s{\rm a}}}\;,
\end{equation}
where $\lambda_b$, $c_1$ and $N_{s{\rm a}}$ can be read from (17) and (18).
(For examble, in the case of the $\Lambda$ Baryons we have $c_1=0$,
$\lambda_b=\lambda_u\lambda_d\lambda_s$ and $N_{s{\rm a}}=1$.)
Therefore
\begin{equation}
\lambda_i\frac{\partial\lambda_{\rm a}}{\partial\lambda_i}=
N_{i{\rm a}}\lambda_b\{\lambda\}\gamma_s^{N_{s{\rm a}}}\;,\;i=1,2,3\;
\end{equation}
and
\begin{equation}
\lambda_i\frac{\partial\lambda_{\rm a}}{\partial\lambda_i}=
N_{s{\rm a}}\lambda_b\{\lambda\}\gamma_s^{N_{s{\rm a}}}\;,\;i=4\;.
\end{equation}
Given the above set of equations there will be a corresponding family of
antiparticles for which
\begin{equation}
\lambda_i\frac{\partial\lambda_{\rm a}}{\partial\lambda_i}=
-N_{i{\rm a}}\lambda_b\{\lambda\}^{-1}\gamma_s^{N_{s{\rm a}}}\;,\;i=1,2,3\;
\end{equation}
and
\begin{equation}
\lambda_i\frac{\partial\lambda_{\rm a}}{\partial\lambda_i}=
N_{s{\rm a}}\lambda_b\{\lambda\}^{-1}\gamma_s^{N_{s{\rm a}}}\;,\;i=4\;.
\end{equation}
will hold true.

The last four equations applied to (32) give
\begin{equation}
\sum_{\rm a}(\lambda_b\{\lambda\}-\lambda_b\{\lambda\}^{-1})
N_{i{\rm a}}\gamma_s^{N_{s{\rm a}}}F_{\rm a}(V,T)=0\;,\;i=1,2,3\;,
\end{equation}
\begin{equation}
\sum_{\rm a}(\lambda_b\{\lambda\}+\lambda_b\{\lambda\}^{-1})
N_{s{\rm a}}\gamma_s^{N_{s{\rm a}}}F_{\rm a}(V,T)=0\;,\;i=4\;,
\end{equation}
where the index ``${\rm a}$'' runs solely over particles. Next we see that
the equation (39) is not possible to hold true because the left
part is always positive (the numbers $N_{s{\rm a}}$ are physical). Therefore
an extremum of the temperature with respect to $\gamma_s$ does not exist.
Turning to (38) we observe that a solution could be found if for all
$\{\lambda\}$ we had
\begin{equation}
\lambda_b\{\lambda\}-\lambda_b\{\lambda\}^{-1}=0,\;\forall b
\Leftrightarrow \lambda_b\{\lambda\}=1,\;\forall b\;.
\end{equation}
An obvious solution for (40) is
\begin{equation}
\lambda_u=\lambda_d=\lambda_s=1 \Leftrightarrow \mu_u=\mu_d=\mu_s=1\;.
\end{equation}
The last equation defines an extremum for the critical temperature with
constant $\gamma_s$.

On the other hand we have
\begin{equation}
\frac{\partial}{\partial \lambda_j}\left(\lambda_i
\frac{\partial \varphi}{\partial \lambda_i}\right)=\frac{1}{\lambda_j}
\sum_{\rm a}(\lambda_b\{\lambda\}+\lambda_b\{\lambda\}^{-1})
N_{i{\rm a}}N_{j{\rm a}}\gamma_s^{N_{s{\rm a}}}F_{\rm a}(V,T)>0\;,\;i=1,2,3\;.
\end{equation}
That is, for every value of $i=1,2,3$ each one of the above equations, once two
values among the $\{\lambda\}$ are fixed, will have a unique solution. That
happens, because from (42), one can infer that
$\lambda_i\frac{\partial \varphi}{\partial \lambda_i}$ is a genuine rising
function with respect to $\lambda_j$ and so it has a unique solution. By
extension the simultaneous solution of the three equations will be unique.
So the extremum we have calculated is unique. This extremum cannot
correspond to minimum, since the critical surface has zero critical
temperature for non zero chemical potentials (e.g. see Figs. 1,3). Since always
$T\geq 0$, if the point which corresponds to (41) was a local minimum, then
we should have another extremum somewhere else, which is imposible, since
the extremum is unique. Therefore (41) corresponds to a {\it total} maximum
for a given value of $\gamma_s$.

\vspace{2cm}
{\Large \bf Appendix B}

\vspace{0.1cm}
In our study we have to calculate $r$ constraints ($r=1$ for isospin
symmetry and $r=2$ for isospin non-symmetry) and different particle
multiplicities as functions of the thermodynamical variables
$(T,\{\lambda\})$. In general all these quatities can be written as
\begin{equation} 
R_j(T,\{\lambda\})\equiv
\int_0^T \frac{1}{y^5} \frac{Q_j(y,\{\lambda\})}{2-\exp [G(y,\{\lambda\})]}
dy\;, 
\end{equation} 
where
\begin{equation} 
R_j\equiv H_j\;,\;j\leq r\;\;\;R_j\equiv N_{j-r}^{theory}\;,\;j > r\;,
\end{equation} 
where $H$ and $N^{theory}$ are given from (23) and (27), respectively, and
\begin{equation} 
Q_j\equiv F_j\;,\;j\leq r\;\;\;,\;\;\;
Q_j\equiv \frac{VT^3}{4\pi^3 H_0}\left.\left[
\frac{\partial\varphi(y,\{\lambda\},\cdots,\lambda_{j-r},\cdots)}
{\partial\lambda_{j-r}}\right]\right|_{\cdots=\lambda_{j-r}=\cdots=1}
\;,\;j > r \;,
\end{equation}
with $F_j$ given from (24) and (25).

In order to evaluate the optimized set of variables $(T,\{\lambda\})$ in our
large working space we have to turn to the use of the generalized
Newton-Raphson method which converges quickly but requires the knowledge of the
derivatives of (43). If we try to calculate these derivatives with respect to
a fugacity $\lambda_i$ from (43) we find
\[
\left.\frac{\partial R_j(T,\{\lambda\})}{\partial \lambda_i}\right|_T=
\int_0^T\frac{dy}{y^5}\left\{
\frac{\exp [G(y,\{\lambda\})]} 
{\{2-\exp [G(y,\{\lambda\}]\}^3} 
\frac{\partial \varphi (y,\{\lambda\})}{\partial \lambda_i} 
Q_j(y,\{\lambda\})\right.+\]
\begin{equation}
\hspace{6cm}\left.\frac{1}{2-\exp [G(y,\{\lambda\})]}
\frac{\partial Q_j(y,\{\lambda\})}{\partial \lambda_i} 
\right\}. 
\end{equation} 

With the use of the above equation the Newton-Raphson method can proceed
for all points of the hadronic space which are not close to the critical
surface. Problems, however, are encountered when the quantities (46) have to be
evaluated {\it near} and even more {\it on} the critiacal surface.
When $T\rightarrow T_{cr}$ the function to be integrated in (46) contains a
non-integrable singularity of the form $(2-\exp[G])^{-3}$. This singularity
cannot be integrated even if we use the variable
$z=2-\exp [G(y,\{\lambda\})]$, as we did in Refs [7,8].

On the other hand the quantities $R_j$ can be expressed as functions of a new
set of variables. This new set can be formed if we replace the temperature T
in favour of the function $\varphi$, or equivalently the $z$ variable. Then
the $R_j$ can be given from
\begin{equation} 
R_j(z,\{\lambda\})=\int_1^{\tilde{z}}\frac{d\tilde{z}}{\tilde{z}-2}\cdot\left[ 
\frac{\textstyle Q_j (y,\{\lambda\})} 
y^5 \cdot\frac{\textstyle \partial \varphi (y,\{\lambda\})} 
{\textstyle \partial y} \right]_ 
{\textstyle \tilde{z}=2-\exp[G(y,\{\lambda\})]}\;\;. 
\end{equation} 
To be able to proceed with the Newton-Raphson method, in this case,
we calculate instead of (46), the derivatives of
$R_j$ when $z$ is constant, i.e. derivatives of the form
$\left. \frac{\partial R_j(z,\{\lambda\})}{\partial \lambda_i}\right|_
{\textstyle z}$. These derivatives should not present any singularity for any
value of $z$ (even for $z=0$ when we are on the critical surface) because,
as can be seen from (47), the $R_j$ can be evaluated for any values of $z$ and
$\{\lambda\}$.

In order to proceed with the evaluation of the derivatives of (47) we can
assume that we are standing on a surface of constant $\varphi$ or
equivalently constant $z$. We then let the variation to the fugacity
$\lambda_i$ be $d\lambda_i$ without ever leaving the above mentioned surface.
Then the variation in $z$ is
\begin{equation} 
dz=\frac{dz}{d\lambda_i}d\lambda_i=0
\end{equation}
Since using the BE we have
\begin{equation} 
\frac{dG}{d\varphi}=\frac{1}{2-e^G}\;,
\end{equation}
we arrive at
\begin{equation} 
\frac{dz}{d\lambda_i}=\frac{dz}{dG}\frac{dG}{d\varphi}
\left.\frac{\partial\varphi}{\partial\lambda_i}\right|_z=
\frac{-e^G}{2-e^G}
\left.\frac{\partial\varphi}{\partial\lambda_i}\right|_z\;.
\end{equation}
From the last two equations we conclude that
\begin{equation}
\left.\frac{\partial\varphi(z,\{\lambda\})}{\partial\lambda_i}\right|_z =0\;.
\end{equation}
Let us comment that on the critical surface $2-e^G=0$, so again equation (51)
is to hold if equations (48) and (50) are to be fulfilled.
If we then express the $z$ variable as function of the temperature $y$ and
the fugacities, $z=z(y,\{\lambda\})$, then equation (51) leads to
\begin{equation}
\frac{\partial\varphi(y,\{\lambda\})}{\partial y} \cdot 
\left.\frac{\partial y}{\partial \lambda_i}\right|_z +
\frac{\partial\varphi(y,\{\lambda\})}{\partial\lambda_i}=0 
\Rightarrow 
\left.\frac{\partial y}{\partial \lambda_i}\right|_z =
-\frac{\partial\varphi / \partial\lambda_i}{\partial\varphi / \partial y}
\end{equation}
The last relation shows us how temperature is varied with the fugacity
$\lambda_i$ on a surface of constant $z$.

Using the definition
\begin{equation}
V_j(y,\{\lambda\})\equiv
\frac{\textstyle Q_j (y,\{\lambda\})}
{y^5 \cdot\frac{\textstyle \partial \varphi (y,\{\lambda\})}
{\textstyle \partial y}}\;,
\end{equation}
the derivatives we seek can be expressed as
\begin{equation} 
\left. \frac{\partial R_j(z,\{\lambda\})}{\partial \lambda_i}\right|_
{\textstyle z}=\int_1^z\frac{d \tilde{z}}{\tilde{z}-2}\cdot\left[
\frac{\textstyle V_j [y(\tilde{z},\{\lambda\}),\{\lambda\}]} 
{\textstyle \partial \lambda_i}\right]\;. 
\end{equation} 

But
\[
\frac{\textstyle V_j [y(\tilde{z},\{\lambda\}),\{\lambda\}]} 
{\textstyle \partial \lambda_i}= 
\frac{\textstyle V_j [y,\{\lambda\}]}{\textstyle y} 
\left.\frac{\textstyle y}{\textstyle \lambda_i}\right|_{\textstyle z}+
\frac{\textstyle V_j [y,\{\lambda\}]}{\textstyle \partial \lambda_i}= 
\hspace{5cm}
\]
\[
=\left[\left(y^5\frac{\textstyle \partial \varphi}{\textstyle \partial y}
\right)^{-1}\frac{\textstyle \partial Q_j}{\textstyle \partial y}-
\frac{\textstyle 5}{\textstyle y^6}
\left(\frac{\textstyle \partial \varphi}{\textstyle \partial y}\right)^{-1}
Q_j-
\frac{\textstyle Q_j}{\textstyle y^5}
\left(\frac{\textstyle \partial \varphi}{\textstyle \partial y}\right)^{-2}
\frac{\textstyle \partial^2 \varphi}{\textstyle \partial y^2}\right]\cdot
\left(-\frac{\textstyle \partial \varphi/\partial\lambda_i}
{\textstyle \partial \varphi/y}\right)+
\hspace{1cm}\]
\[\hspace{5cm}
+\left(y^5\frac{\textstyle \partial \varphi}{\textstyle \partial y}
\right)^{-1}\frac{\textstyle \partial Q_j}{\textstyle \partial \lambda_i}-
\frac{\textstyle Q_j}{\textstyle y^5}
\left(\frac{\textstyle \partial \varphi}{\textstyle \partial y}\right)^{-2}
\frac{\textstyle \partial^2 \varphi}{\textstyle \partial y\partial \lambda_i}=
\]
\begin{equation}
=y^{-5}\left(\frac {\partial \varphi}{\partial y}\right)^{-2}\left[
\frac{\partial \varphi}{\partial y} \frac{\partial Q_j}{\partial \lambda_i}-
\frac{\partial \varphi}{\partial \lambda_i} \frac{\partial Q_j}{\partial y}+
Q_j\left(-\frac{\partial^2 \varphi}{\partial y \partial \lambda_i}+
\frac{\partial^2 \varphi}{\partial y^2}
\frac{\partial\varphi / \partial\lambda_i}{\partial\varphi / \partial y}
+\frac{5}{y}\frac{\partial \varphi}{\partial \lambda_i}\right)\right]. 
\end{equation}
From the last two equations we conclude that
\[
\left.\frac{\textstyle \partial R_j(z,\{\lambda\})}
{\textstyle \partial \lambda_i}\right|_{\textstyle z}=
\int_1^z\frac{d\tilde{z}}{tilde{z}-2}\left\{y^{-5}
\left(\frac {\partial \varphi}{\partial y}\right)^{-2}\left[
\frac{\partial \varphi}{\partial y} \frac{\partial Q_j}{\partial \lambda_i}-
\frac{\partial \varphi}{\partial \lambda_i} \frac{\partial Q_j}{\partial y}+
\right.\right.
\]
\begin{equation}
\left.\left.
Q_j\left(-\frac{\partial^2 \varphi}{\partial y \partial \lambda_i}+
\frac{\partial^2 \varphi}{\partial y^2}
\frac{\partial\varphi / \partial\lambda_i}{\partial\varphi / \partial y}
+\frac{5}{y}\frac{\partial \varphi}{\partial \lambda_i}\right)\right]
\right\}_{\textstyle \tilde{z}=2-\exp [G(y,\{\lambda\})]}. 
\end{equation}

As it is known the function to be $z$-intergrated has not so good behaviour
near $z=1$. So it is better to break the above integral in two parts:

\[\left.\frac{\textstyle \partial R_j(z,\{\lambda\})}
{\textstyle \partial \lambda_i}\right|_{\textstyle z}=
\hspace{11cm}\]

\[\int_0^{T_1}\frac{dy}{2-e^G}y^{-5}
\left(\frac {\partial \varphi}{\partial y}\right)^{-1}\left[
\frac{\partial \varphi}{\partial y} \frac{\partial Q_j}{\partial \lambda_i}-
\frac{\partial \varphi}{\partial \lambda_i} \frac{\partial Q_j}{\partial y}+
Q_j\left(\frac{\partial^2 \varphi}{\partial y \partial \lambda_i}-
\frac{\partial^2 \varphi}{\partial y^2}
\frac{\partial\varphi / \partial\lambda_i}{\partial\varphi / \partial y}
+\frac{5}{y}\frac{\partial \varphi}{\partial \lambda_i}\right)\right]+ \]

\[\int_{z_1}^z\frac{d\tilde{z}}{\tilde{z}-2}\left\{y^{-5}
\left(\frac {\partial \varphi}{\partial y}\right)^{-2}\left[
\frac{\partial \varphi}{\partial y} \frac{\partial Q_j}{\partial \lambda_i}-
\frac{\partial \varphi}{\partial \lambda_i} \frac{\partial Q_j}{\partial y}+
\right.\right.\hspace{7cm}\]

\begin{equation}
\hspace{4cm}\left.\left.
Q_j\left(\frac{\partial^2 \varphi}{\partial y \partial \lambda_i}-
\frac{\partial^2 \varphi}{\partial y^2}
\frac{\partial\varphi / \partial\lambda_i}{\partial\varphi / \partial y}
+\frac{5}{y}\frac{\partial \varphi}{\partial \lambda_i}\right)\right]
\right\}_{\textstyle \tilde{z}=2-\exp [G(y,\{\lambda\})]}. 
\end{equation}
In the above relation $z_1=2-\exp [G(T_1,\{\lambda\})$ and a
good choice is $z_1=0.5$. If $z>0.5$ we are not close to the critical surface
and the second integral does not have to be calculated. So in general we can
set $z_1=max\{z,0.5\}$.

The derivatives of $R_j$ with respect to $z$ can be calculated easily. They
simply read
\begin{equation} 
\left. \frac{\partial R_j(z,\{\lambda\})}{\partial z}\right|_
{\textstyle \lambda_i}=
\frac{1}{z-2}\cdot y^{-5} \cdot \left\{
\frac{\textstyle \partial \varphi [y(z,\{\lambda\}),\{\lambda\}]}
{\textstyle \partial y}\right\}^{-1}
{\textstyle Q_j [y(z,\{\lambda\}),\{\lambda\}]}\;.
\end{equation} 

With the above relations the minimization of the $\chi^2$ function can
proceed with the use of the Newton-Raphson method. Relation (58) 
can also be used to find out whether the absolute minimum
of $\chi^2$ is outside or inside the critical surface.
Suppose we locate the minimum value of $\chi^2=(\chi^2)_1$ {\it on} the
critical surface and this value corresponds to the point
$(z,\{\lambda\})=(0,\{\lambda^0\})$. Then the absolute minimum of $\chi^2$ is
located {\it inside} the hadronic phase if for this point we have
\begin{equation} 
\left. \frac{\partial \chi^2 (z,\{\lambda^0\})}{\partial z}\right|_
{\textstyle z=0}<0\;.
\end{equation} 
If the above relation is not fulfilled the absolute minimum of $\chi^2$ lies
on the {\it outside}.

An alternative method to verify the same thing consists of locating the
minimum value of $\chi^2$ on a surface near the critical one inside the
hadronic phase. Let this value be $(\chi^2)_2$. The absolute minimum of
$\chi^2$ is located {\it inside} the hadronic phase if
$(\chi^2)_2<(\chi^2)_1$ and outside otherwise. For the two fits we have
performed for $S+Ag$ we had to process 192 points. All these points have
given us the same results with the use of the two methods.

\newpage
{\large{\bf Table Captions}} 
\newtheorem{f}{Table} 
\begin{f} 
\rm The fitted parameters and the $\chi^2/dof$ values for different
fits through SSBM in the experimentally measured full phase space
multiplicities in the $S+Ag$ interaction. In the first fit all the
multiplicities are included while in the following fits we exclude each time one
mutiplicity. $T_0$ is set to 190 MeV.
\end{f} 
\begin{f} 
\rm Experimentally measured full phase space multiplicities in the
$S+Ag$ interaction and their theoretically fitted values by the SSBM, 
with the inclusion of the $h^-$ multiplicity and without it
(cases A: $T_0=182.94$ MeV and B: $T_0=183.5$ MeV).
\end{f}
\begin{f} 
\rm Calculation of the correction factor 
$f_i=\frac{(N_{IHG-BF})_i-(N_{IHG-BO})_i}{(N_{IHG-BO})_i}$ for the $i$th
particle species measured in $4\pi$ phase space in the $S+Ag$ interaction.
For the calculation of $f_i$ the IHG formalism has been used, while the
thermodynamical variables have been extracted from the SSBM fit with
$h^-$ and without $h^-$ (cases A and B).
\end{f}
\begin{f} 
\rm Results of the analysis by SSBM of the experimental
data from the $S+Ag$ interaction ($4\pi$ phase space), with the inclusion
of the $h^-$ multiplicity and without it (cases A and B). 
\end{f}
\begin{f}
\rm Particle ratios from the experimentally measured full phase space
multiplicities for the $S+Ag$ interaction used in the analysis with $h^-$.
\end{f}

\newpage
\vspace{2cm}
$\;$
\vspace{4cm}
\begin{center}
{\bf $\bf S+Ag$ (NA35) Full phase space}
\end{center}
\begin{center}
\begin{tabular}{|c|ccccccc|} \hline
Excluded & $T$ (MeV) & $\lambda_u$ & $\lambda_d$ & $\lambda_s$ & $\gamma_s$ &
$VT^3/4\pi^3$ & $\chi^2/dof$ \\ \hline\hline 
none & 170.943 & 1.540 & 1.582 &  1.088 &  0.662 & 2.883 & 10.35/3 \\
${K_s}^0$ & 162.411 & 1.551 & 1.588 & 1.132 & 0.749 & 3.869 & 4.72/2 \\
$\Lambda$ & 170.363 & 1.523 & 1.563 & 1.093 & 0.616 & 3.063 & 9.41/2 \\ 
$\overline{\Lambda}$ & 158.361 & 1.625 & 1.665 & 1.174 & 0.636 & 4.239 & 7.58/2 \\ 
$\overline{p}$ & 175.180 & 1.482 & 1.520 & 1.067 & 0.596 & 2.807 & 4.55/2 \\ 
$p-\overline{p}$ & 170.626 & 1.538 & 1.579 & 1.090 &  0.665 & 2.904 & 10.33/2 \\ 
$B-\overline{B}$ & 172.040 & 1.554 & 1.597 & 1.083 &  0.656 & 2.786 & 8.52/2 \\ 
$h^-$ & 180.779 & 1.642 & 1.702 & 1.011 & 0.839 & 1.261 & 1.64/2 \\ \hline 
\end{tabular}
\end{center} 
\begin{center}
Table 1.
\end{center}

\newpage
\vspace{2cm}
\begin{center}
{\bf $\bf S+Ag$ (NA35) Full phase space} 
\end{center}

\begin{center}
\begin{tabular}{|c|cccc|} \hline 
Particles & Experimental & Calculated & Calculated   & Calculated   \\
          & Data         & with $h^-$ & without $h^-$& without $h^-$\\
          &              &            & (Case A)     & (Case B)     \\ \hline\hline
${K_s}^0$ & $15.5\pm 1.5$ & 17.613 & 15.181 & 15.155 \\
$\Lambda$ & $15.2\pm 1.2$ & 14.490 & 15.424 & 15.429 \\ 
$\overline{\Lambda}$ & $2.6\pm 0.3$ & 2.3998 & 2.5502 & 2.5538 \\ 
$\overline{p}$ & $2.0\pm 0.8$ & 3.4614 & 2.3612 & 2.3547 \\ 
$p-\overline{p}$ & $43\pm 3$ & 42.600 & 40.931 & 40.937 \\ 
$B-\overline{B}$ & $90\pm 10$ & 101.39 & 99.307 & 99.325 \\ 
$h^-$ & $186\pm 11$ & 170.68$^{\rm a}$ & 128.85$^b$ & 128.57$^c$ \\ \hline 
\end{tabular} 
\end{center}

{\footnotesize
$^{\rm a}$ A correction factor 1.0236 has been included for the effect of 
Bose statistics. 
 
$^b$ A correction factor 1.0190 has been included for the effect of Bose 
statistics. This multiplicity is not included in the fit.

$^c$ A correction factor 1.0188 has been included for the effect of Bose 
statistics. This multiplicity is not included in the fit.}
 
\begin{center}
Table 2.
\end{center}

\vspace{0.5cm}
\begin{center}
{\bf $\bf S+Ag$ (NA35) Full phase space} 
\end{center}
\begin{center}
\begin{tabular}{|c|ccc|} \hline 
Particles ($i$)&$f_i$($\%$) for the&$f_i$($\%$) for the   &$f_i$($\%$) for the \\
               &fit with $h^-$      &fit without $h^-$ (A)&fit without $h^-$ (B)\\ \hline\hline
${K_s}^0$ & $0.462$ & $0.641$ & $0.645$ \\
$\Lambda$ & $0.053$ & $0.232$ & $0.237$ \\ 
$\overline{\Lambda}$ & $-0.142$ & $-0.410$ & $-0.418$ \\ 
$\overline{p}$ & $0.146$ & $0.105$ & $0.105$ \\ 
$p-\overline{p}$ & $-0.388$ & $-0.431$ & $-0.432$ \\ 
$B-\overline{B}$ & $-0.434$ & $-0.417$ & $-0.417$ \\ 
$h^-$ & $2.357$ & $1.897$ & $1.877$ \\ \hline 
\end{tabular} 
\end{center} 
 \begin{center}
Table 3.
\end {center}

\newpage
\vspace{2cm}
\begin{center}
{\bf $\bf S+Ag$ (NA35) Full phase space} 
\end{center}
\begin{center}
\begin{tabular}{|c|ccc|} \hline
Fitted     & Fitted with $h^-$ & Fitted without $h^-$ & Fitted without $h^-$ \\ 
Parameters &                   & (Case A) & (Case B)  \\ \hline\hline
$T$ (MeV) & $170.6\pm 5.9$ & 176.3 & $176.8\pm2.1$ \\
$\lambda_u$ & $1.544\pm 0.046$ & 1.640 & $1.641\pm0.074$ \\ 
$\lambda_d$ & $1.586\pm 0.050$ & 1.700 & $1.701\pm0.084$ \\ 
$\lambda_s$ & $1.084\pm 0.036$ & 1.012 & $1.011\pm0.047$ \\
$\gamma_s$ & $0.670\pm 0.073$ & 0.836 & $0.84\pm0.12$ \\
$VT^3/4\pi^3$ & $2.74\pm 0.71$ & 1.23 & $1.22\pm0.60$ \\ 
$\chi^2/dof$ & $9.37\;/\;3$ & $1.654\;/\;2^{\;e}$ & $1.652\;/\;2$ \\ 
$\mu_u$ (MeV) & $74.1\pm5.7$ & 87.3 & $87.6\pm8.0$ \\ 
$\mu_d$ (MeV) & $78.7\pm6.0$ & 93.5 & $93.9\pm8.8$ \\ 
$\mu_s$ (MeV) & $13.8\pm5.7$ & 2.2 & $2.0\pm8.1$ \\
$P_{INSIDE}$ & $100\%\;(128/128)$ & $48.44\%\;(31/64)$ & $-$ \\ \hline 
\end{tabular} 
\end{center} 

{\footnotesize
$^e$ It is the minimum of $\chi^2$ within the Hadron Gas with $T_0=183$ 
MeV (for $\gamma_s=1$), not the absolute minimum.}

\begin{center}
Table 4.
\end {center}

\vspace{0.5cm}
\begin{center}
{\bf $\bf S+Ag$ (NA35) Full phase space} 
\end{center} 
\begin{center}
\begin{tabular}{|c|c|} \hline
Particle ratios for& Experimental \\ 
the fit with $h^-$ & Values       \\ \hline\hline
$K^0_s/h^-$& $0.0833\pm 0.0095$ \\
$\Lambda/h^-$& $0.0817\pm 0.0081$ \\
$\overline{\Lambda}/h^-$& $0.0140\pm 0.0018$ \\
$\overline{p}/h^-$& $0.0108\pm 0.0043$ \\
$p-\overline{p}/h^-$& $0.231\pm 0.021$ \\
$B-\overline{B}/h^-$& $0.484\pm 0.061$ \\ \hline
\end{tabular} 
\end{center} 
\begin{center}
Table 5.
\end{center}

\newpage
\vspace{2cm}
{\large{\bf Figure Captions}}
\newtheorem{g}{Figure} 
\begin{g}
\rm Projections on the $(\mu_u,T)$-plane of intersections of constant $\mu_d$
and constant $\mu_s$ of the critical surface
$\varphi(T,\mu_u,\mu_d,\mu_s,\gamma_s)=\ln 4-1$ for $T_0=183$ MeV at
$\gamma_s=1$.
\end{g}
\begin{g}
\rm Projections on the $(\mu_d,\mu_u)$-plane of intersections of constant $T$
and constant $\mu_s$ of the critical surface
$\varphi(T,\mu_u,\mu_d,\mu_s,\gamma_s)=\ln 4-1$ for $T_0=183$ MeV at
$\gamma_s=1$.
\end{g}
\begin{g}
\rm Projections on the $(T,\mu_s)$-plane of intersections of constant $\mu_u$
and constant $\mu_d$ of the critical surface
$\varphi(T,\mu_u,\mu_d,\mu_s,\gamma_s)=\ln 4-1$ for $T_0=183$ MeV at
$\gamma_s=1$.
\end{g}
\begin{g}
\rm Projections on the $(T,\mu_s)$-plane of intersections, at fixed
$\lambda_u$ $(\mu_u/T=0.4)$, of the $<S>=0$ and $<B>=2\beta<Q>$ surfaces for
the SSBM and the IHG for different values of $\beta$. For the SSBM case $T_0$ is set
at 183 MeV.
\end{g}
\begin{g}
\rm Projections on the $(\mu_u,\mu_d)$-plane of intersections, at fixed
$\lambda_u$ $(\mu_u/T=0.4)$, of the $<S>=0$ and $<B>=2\beta<Q>$ surfaces for
the SSBM and the IHG, for different values of $\beta$. For the SSBM case $T_0$ is set
at 183 MeV.
\end{g}
\begin{g}
\rm Projections on the $(T,\mu_u)$-plane of the intersection
of the $<S>=0$ and $<B>=2\beta<Q>$ surfaces for the SSBM with the critical
surface for different values of $\beta$. $T_0$ is set at 183 MeV.
\end{g}
\begin{g}
\rm Projections on the $(\mu_u,\mu_d)$-plane of the intersection
of the $<S>=0$ and $<B>=2\beta<Q>$ surfaces for the SSBM with the critical
surface for different values of $\beta$. $T_0$ is set at 183 MeV.
\end{g}
\begin{g}
\rm Projections on the $(\mu_u,\mu_s)$-plane of the intersection
of the $<S>=0$ and $<B>=2\beta<Q>$ surfaces for the SSBM with the critical
surface for different values of $\beta$. $T_0$ is set at 183 MeV.
\end{g}
\begin{g} 
\rm Experimental particle ratios in the ($\mu_u,T$)-plane for the $S+Ag$ 
interaction measured in $4\pi$ phase space with $\gamma_s$ set to 0.67. 
The point and the cross correspond to the $\chi^2$ fit with the $h^-$. The
thick solid line represent the limits of the hadronic phase (HG) as 
set by the SSBM. 
\end{g} 
\begin{g} 
\rm Experimental particle multiplicities in the ($\mu_u,T$)-plane for $S+Ag$ 
interaction measured in $4\pi$ phase space with $\gamma_s$ set to 0.84 and
$VT^3/4\pi^3$ set to 1.23.
The point represented by the solid circle corresponds to the location of the 
least value within the hadron gas of $\chi^2$, 
without the $h^-$. The lines which correspond to $K^0_s=17$, $h^-=175$ and
$h^-=197$ lie outside the hadronic domain, as set by the S-SBM.
\end{g}
\begin{g}
\rm The same diagram as Fig. 10, but with an enlargement of a smaller area
to show the common overlapping region (shaded area) within the hadronic
phase which is compatible with all the measured multiplicities, except $h^-$.
The lines wich corespond to $\Lambda=16.4$, $\bar{\Lambda}=2.9$,
$\bar{p}=1.2$, $\bar{p}=2.8$ and $p-\bar{p}=46$ are outside the region of the
diagram and enclose the shaded region.
\end{g}
\begin{g} 
\rm Comparison between the experimentally measured multiplicities in 
$4\pi$ phase space and the theoretically calculated values in the fit with 
$h^-$ and without $h^-$ (cases A and B) for the $S+Ag$ interaction. The 
difference is measured in units of the relevant experimental error. 
\end{g} 
\begin{g} 
\rm $(\mu_B,T)$-phase diagramme with points obtained from fits to
$p+\bar{p}$ [8], $S+S$ [8] and $S+Ag$ data and corresponding critical curves
given by SSBM.
\end{g}

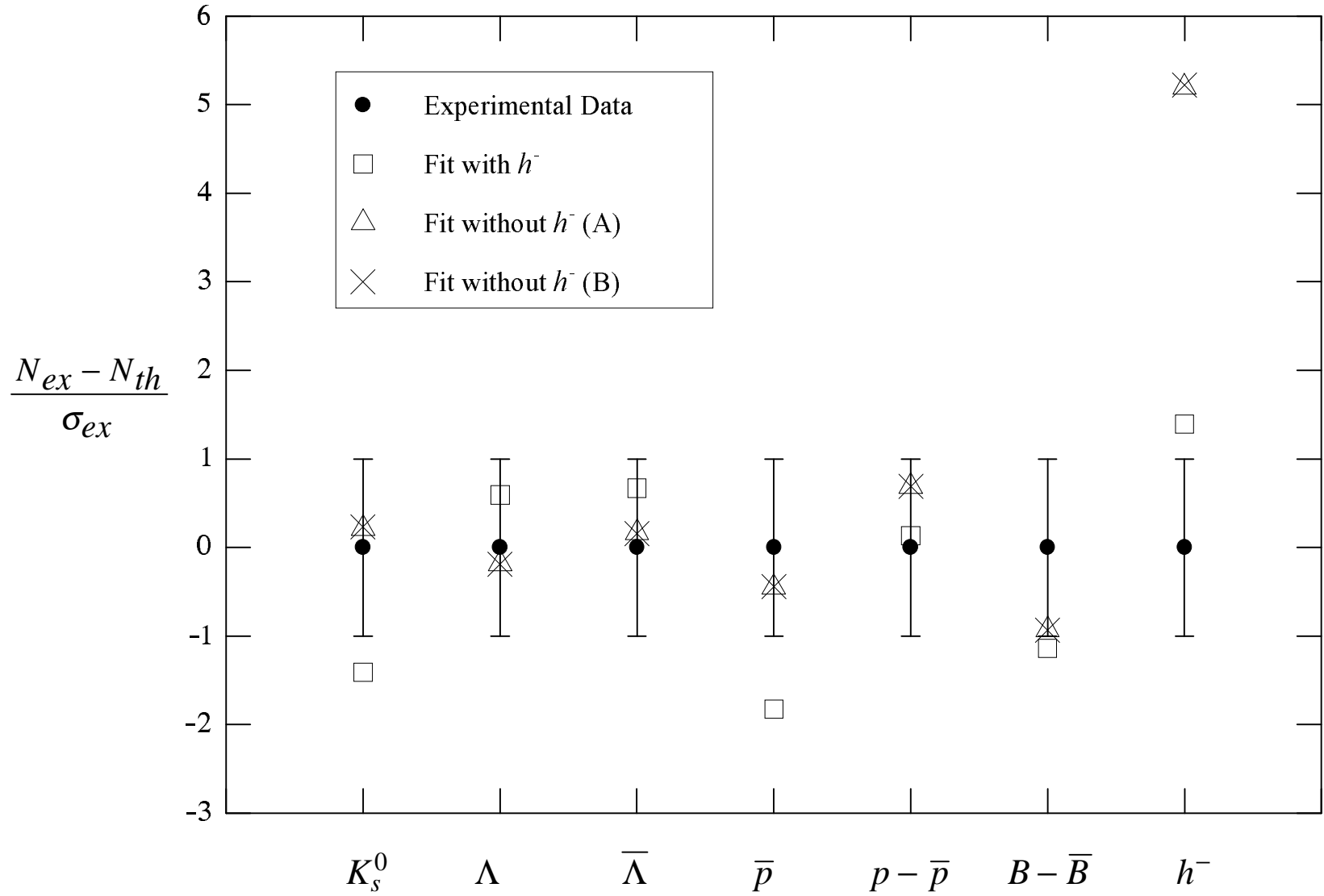
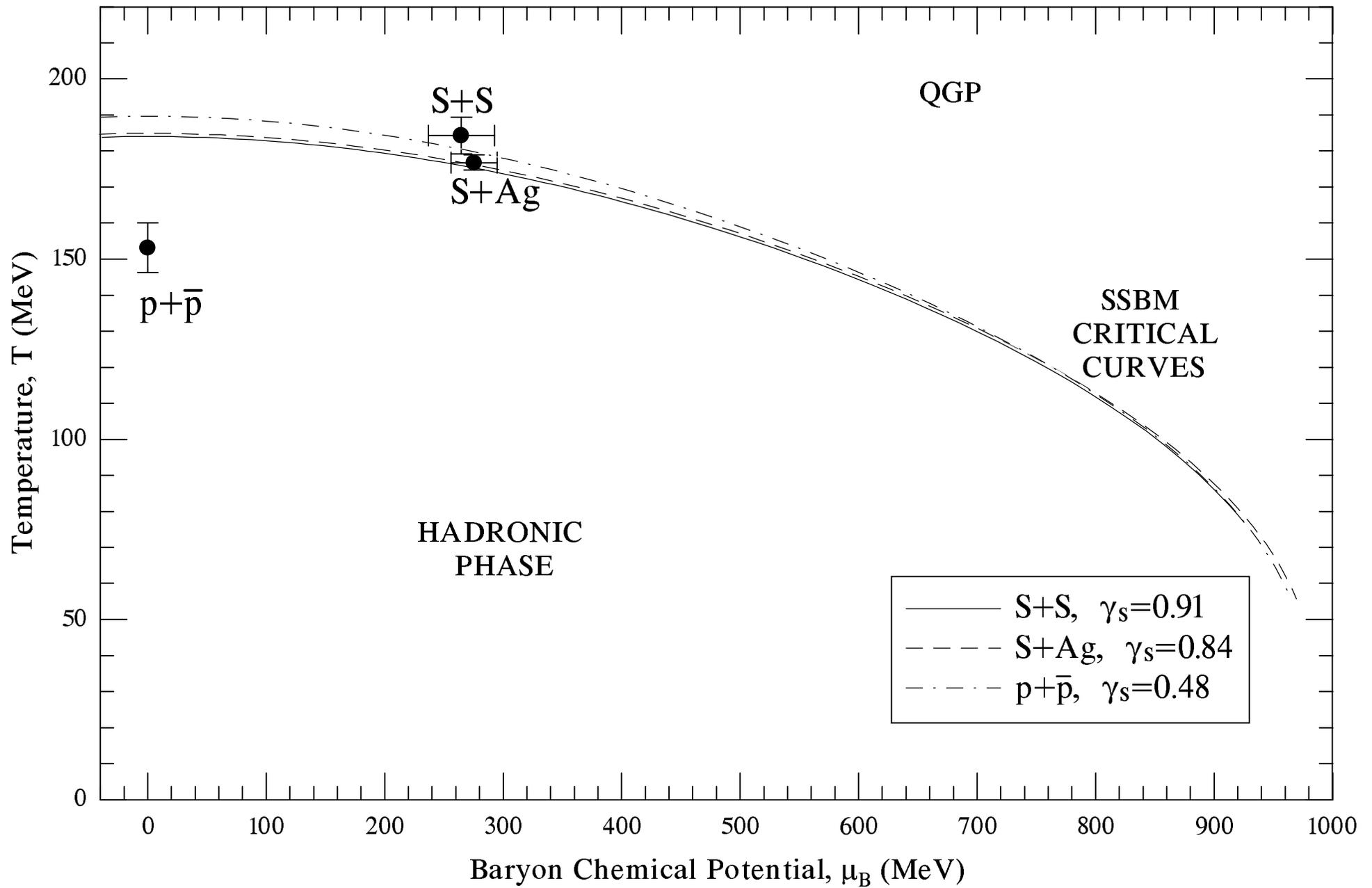
\end{document}